\documentstyle[12pt]{article}
\topmargin--0cm
\oddsidemargin--1mm
\begin{document}
\def\R{\relax{\rm I\kern-.18em R}}
\def\1{\relax{\rm 1\kern-.27em I}}
\def\l{\lambda}
\newcommand{\Z}{Z\!\!\! Z}
\newcommand{\ph}{PS_{ph}}
\def\one{1\hskip-.37em 1}
\def\no{\nonumber}
\def\D{\cal D}
\def\Prod{\prod}
\def\ir{{\rm I}\hskip-.2em{\rm R}}
\def\E{{\rm E}\hskip-.55em{\rm I}\hskip.2em}
\def\half{\textstyle{\frac{1}{2}}}
\def\iN{{\rm I}\hskip-.2em{\rm N}}
\def\irtwo{{\rm I}\hskip-.2em{\rm R}^2}
\def\irn{{\rm I}\hskip-.2em{\rm R}^n}
\def\iH{{\rm I}\hskip-.2em{\rm H}}
\def\ra{\rightarrow}
\def\tint{{\textstyle\int}}
\def\d{\partial}
\def\o{\overline}
\def\b{\begin{eqnarray*}}     
\def\e{\end{eqnarray*}}       
\def\bn{\begin{eqnarray}}     
\def\en{\end{eqnarray}}       
\def\<{\langle}
\def\>{\rangle}
\def\{{\lbrace}
\def\}{\rbrace}
\def\D{{\cal D}}
\def\H{{\cal H}}
\bibliographystyle{unsrt}

\begin{center}
{\LARGE  An Introduction to\\    
Coordinate-free Quantization and its\\
Application to Constrained Systems}\footnote{To appear 
in the Proceedings of the 2nd Jagna Workshop, 
Jagna, Bohol, Philippines, January, 1998.}

\vskip 0.3cm
John R. KLAUDER $ ^a$\ \ \ and\ \ \ Sergei V. SHABANOV
$ ^{b,}$\footnote{On leave from Laboratory of Theoretical
Physics, JINR, Dubna, Russia.}

\vskip 0.3cm
$ ^a${\em Departments of Physics and Mathematics,
 University of Florida,
Gainesville FL-32611, USA }\\
$ ^b${\em Institute for Theoretical Physics, FU-Berlin,
Arnimallee 14, D-14195, Berlin, Germany }
\end{center}

\begin{abstract}
Canonical quantization entails using Cartesian coordinates, 
and Cartesian coordinates exist only in flat spaces. 
This situation can either be questioned or accepted. 
In this paper we offer a brief and introductory overview of how a 
flat phase space metric can be incorporated into a covariant, 
coordinate-free quantization procedure involving a 
continuous-time (Wiener measure) regularization of 
traditional phase space path integrals. 
Additionally we show how such procedures can be extended 
to incorporate systems with constraints and illustrate 
that extension for special systems.

\end{abstract}
\subsection*{Introduction}

In order to quantize a system with constraints it is of course first 
necessary to have a quantization procedure for systems without
constraints. 
Although the quantization of systems without constraints 
would seem to be well in hand due to the pioneering work of 
Heisenberg, Schr\"odinger, and Feynman, 
it is a less appreciated fact that all of the standard 
methods of quantization are consistent only in Cartesian 
coordinates \cite{dir}. As a consequence it follows that 
the usual quantization procedures depend---or at 
least seem to depend---on choosing the right set of coordinates 
before promoting $c$-numbers to $q$-numbers. 
This circumstance gives rise to an apparently unwanted coordinate 
dependence on the very process of quantization. 
For systems without constraints this is generally not a major 
problem because an underlying Euclidean space expressed in terms of 
Cartesian coordinates can generally be assumed. However, 
for systems with constraints, the configuration space---let 
alone the frequently more complicated phase space---are generally 
incompatible with a flat Euclidean structure needed to carry 
Cartesian coordinates. Hence, before we can properly quantize systems
with constraints it will be necessary for us to revisit 
the quantization of systems without constraints in order to present a 
coordinate-free procedure for such cases. Only then will we be able to
undertake the program represented by the title to this contribution. 

All individuals engaged in quantization have a natural inclination to
seek a quantization procedure that is as coordinate independent as
possible, and in particular, does not depend on using Cartesian
coordinates inasmuch as the use of such coordinates seems to
contradict the ultimate goal of a coordinate-free approach. When faced
with the need for quantizing in Cartesian coordinates a number of
workers currently seek alternative quantization schemes which avoid 
completely any reference to Cartesian coordinates.  The schemes of
geometric quantization and of deformation quantization, among possibly
other approaches as well, fit into the category of efforts to
eliminate the central role played by Cartesian coordinates, and indeed
to construct a fully coordinate independent formulation. It is
entirely natural to presume that quantization should be coordinate 
independent, and so this approach is most reasonable. 
While these disciplines can be considered and analyzed from a 
mathematical viewpoint without any conceptual difficulty, 
it is not a priori evident that just because these methods have 
the word ``quantization'' in their name that they have an automatic 
connection with physics. Indeed, it may be argued that this is not 
always the case, and this conclusion pertains to the fact that the result does
not in general agree with the results of ordinary quantization in 
the physical sense of the term. Thus such contemporary methods are 
acceptable as mathematical exercises but should not be taken
necessarily as leading to a coordinate-free formulation of 
quantization as it is needed and used in physics.

Such a circumstance naturally leads to the question: Can we find a 
coordinate-free form of quantization that does agree with physics, 
i.e., as a test case to quantize the anharmonic oscillator in accord 
with the usual quantum mechanical result as obtained, say, from the 
Schr\"odinger prescription? The answer in our opinion is yes, and 
the first part of this article is devoted to a brief review of that 
subject. The extension of that procedure to systems with constraints 
is currently in progress, and a preliminary account of part of that
work forms the second part of this article.
 
\section*{Coordinate-free Quantization}
In 1948 Feynman  proposed the path integral in Lagrangian form, and in 1951
he extended the path integral to a phase space form. It is generally 
acknowledged that the phase space formulation is more widely
applicable than the Lagrangian formulation, and it is the phase space 
version on which we shall focus. In particular, the expression for the 
usual propagator is formally given by
\bn  
 {\cal N}\int\exp\{i\tint [p{\dot q}-h(p,q)]\,dt\}\,\D p\,\D q\;. 
\en
As it stands, however, this expression is formal and needs to be
properly defined. There are several ways to do so. One way involves
what may be referred to as a continuous-time regularization in the 
form \cite{kla1}
\bn  &&\hskip-.5cm K(p'',q'',T;p',q',0)\no\\
&&=\lim_{\nu\ra\infty}{\cal N}
\int\exp\{i\tint[p{\dot q}-h(p,q)]\,dt\}
 \,\exp[-(1/2\nu)\tint({\dot p}^2+{\dot q}^2)\,dt]\,\D
p\,\D q. 
\en 
In this expression we have introduced an additional term in the
integrand which is formally set to unity in the limit $\nu\ra\infty$. 
However, that extra factor in the integrand serves as a {\it
  regularizing factor}, and this fact can be seen in 
the alternative---and mathematically precise---expression given by
 \bn K(p'',q'',T;p',q',0)=\lim_{\nu\ra\infty}2\pi e^{\nu T/2}
\int\exp\{i\tint[pdq-h(p,q)dt]\}\,d\mu^\nu_W(p,q)\;, 
\en
where $\mu^\nu_W$ denotes a Wiener measure on a flat phase space 
expressed in Cartesian coordinates and in which $\nu$ denotes the 
diffusion constant. In addition, the nature of the regularization 
forces one to pin (i.e., fix) the values of both $p$ {\it and} $q$ 
at the {\it initial and final times}, namely, $p''=p(T),q''=q(T)$ 
and $p'=p(0), q'=q(0)$. This leads to a nontraditional representation 
of the propagator, and as we assert below the very regularization
itself 
leads to a {\it canonical coherent state representation}. 
Observe, moreover, that (3)  is mathematically well-defined and
totally 
ambiguity free. As such, there can be no ambiguity in factor ordering 
within the quantization procedure at this point, and it is significant
that the very regularization chosen has {\it selected}---or perhaps
even better, {\it preselected}---a particular operator ordering, 
namely antinormal ordering. 

More specifically, on the basis of being a positive-definite function,
one may show that 
 \bn  && \hskip-1cm K(p'',q'',T;p',q',0)\equiv\<p'',q''|\,e^{-i\H T}\,
|p',q'\>\;,\\
   &&|p,q\>\equiv e^{-iqP}\,e^{ipQ}\,|0\>\;,\hskip1cm(Q+iP)\,|0\>=0\;,
\hskip1cm\<0|0\>=1\;,\\
   &&[Q,P]=i\one\;,\\
  &&\one=\tint|p,q\>\<p,q|\,dp\,dq/2\pi\;,\\
  &&\H=\tint h(p,q)|p,q\>\<p,q|\,dp\,dq/2\pi\;,  
\en
the last formula being an alternative expression for antinormal ordering.
We do not attempt to prove these remarks here; for that the reader 
may consult the literature  \cite{dakl}. Instead, we limit our 
discussion to an overview of the general scheme. In that line it is 
important to note the behavior of these expressions under a 
{\it canonical coordinate transformation}. In the classical 
theory, we often let 
\bn  
rds=pdq+dF(s,q)  
\en 
symbolize a canonical change of coordinates where the function $F$ 
serves as the ``generator'' of the coordinate change. In the quantum 
theory, as described here, the paths $p(t)$ and $q(t)$ represent
sample paths of a Wiener process, i.e., Brownian motion paths. As such
these paths are continuous but nowhere differentiable, and thus they 
are more singular than the classical path behavior (e.g., $C^2$) for 
which (9) normally holds. The integral $\tint pdq$ appearing in (3) 
is initially undefined due to the distributional nature of the paths 
involved. There are two standard prescriptions to deal with such 
stochastic integrals, one due to It\^o (I), the other due to 
Stratonovich (S) \cite{str}. The two prescriptions may be
characterized by continuum limits of two distinct discretization 
procedures. If $p_l\equiv p(l\epsilon)$ and $q_l\equiv q(l\epsilon)$ 
for $l\in\{0,1,2,3,\ldots\}$, then 
\bn \tint_I \,pdq=\lim_{\epsilon\ra0}\Sigma p_l(q_{l+1}-q_l)\;,
\hskip1cm \tint_S \,pdq=\lim_{\epsilon\ra0}\Sigma \half(p_{l+1}+
p_l)(q_{l+1}-q_l)\;.  
\en
Generally, the results of these two approaches disagree, and it 
is a feature of the It\^o prescription (``nonanticipating'') that 
the rules of ordinary calculus are generally not obeyed. The It\^o 
prescription has other virtues, but they are not of interest to us 
here. Instead, we adopt the Stratonovich (``midpoint'') prescription 
because it possesses the important feature that the ordinary laws of 
the classical calculus do in fact hold for Brownian motion paths. 
In particular, therefore, (9) also holds for the paths that enter 
the regularized form of the phase space path integral, and as a 
consequence, we find that
  \bn  \tint pdq=\tint rds-\tint dF(s,q)\equiv\tint rds+\tint dG(r,s)
=\tint rds+G(r'',s'')-G(r',s')\;. 
\en  
If we couple this relation with the fact that $h$ transforms 
as a {\it scalar}, i.e., namely, that
   \bn 
{\o h}(r,s)\equiv h(p(r,s),q(r,s))=h(p,q)\;, 
\en
then we learn that under a canonical change of coordinates, 
the coherent state propagator becomes
  \bn &&\hskip-1cm{\o K}(r'',s'',T;r',s',0)\equiv 
e^{i[G(r'',s'')-G(r',s')]}\,K(p(r'',s''),q(r'',s''),T;
p(r',s'),q(r',s'),0)\no\\
&&\hskip2cm=\lim_{\nu\ra\infty}2\pi e^{\nu T/2}\int
\exp\{i\tint[rds+dG(r,s)-{\o h}(r,s)]\,dt\}\,d{\o \mu}^\nu_W(r,s)\;.  
\en
Here ${\o\mu}^\nu_W(r,s)$ denotes the Wiener measure on a flat
two-dimensional phase space no longer expressed, in general, in 
Cartesian coordinates but rather in curvilinear coordinates. 
Observe that the form of this expression is exactly that as given 
in the original coordinates apart from the presence of the total 
derivative $dG$, which leads to nothing more than a phase change 
for the coherent states. In particular, based on the positive-definite
nature of the transformed function we may conclude that 
   \bn&&\hskip-1cm {\o K}(r'',s'',T;r',s',0)=\<r'',s''|\,e^{-i\H T}\,
|r',s'\>\;,\\  &&|r,s\>\equiv
e^{-iG(r,s)}\,e^{-iq(r,s)P}\,e^{ip(r,s)Q}\,
|0\>\;,\hskip.6cm(Q+iP)|0\>=0\;,\hskip.6cm\<0|0\>=1\;,\\
  &&\one=\tint|r,s\>\<r,s|\,dr\,ds/2\pi\;,\\
  &&\H=\tint {\o h}(r,s)|r,s\>\<r,s|\,dr\,ds/2\pi\;.  
\en
Observe carefully that the coherent states have {\it not} changed 
under the coordinate transformation, only their {\it names} have
changed. In addition, the operator $\H$ has {\it not} changed, only 
the functional form of the (lower) symbol associated with it has 
changed. Thus we have achieved a completely covariant formulation 
of quantum theory!

As an example of such a coordinate change, we may cite the simple 
case of the harmonic oscillator for which
$h(p,q)=(p^2+q^2)/2$. If we introduce new canonical coordinates 
(action angle variables) according to $r=(p^2+q^2)/2$ and 
$s=\tan^{-1}(q/p)$---namely, where $F(q,s)=-q^2\cot(s)/2$ and 
$G(r,s)=r\cos(s)\sin(s)$---then it follows that 
  \bn  \half(P^2+Q^2+1)=\tint \half(p^2+q^2)\,|p,q\>\<p,q|\,dp\,dq/2\pi
      =\tint r\,|r,s\>\<r,s|\,dr\,ds/2\pi\;,  
\en
which clearly illustrates the fact that although a classical 
coordinate change has been carried out, all quantum operators, 
such as $Q$, $P$, and particularly the Hamiltonian $\H$, have 
remained completely unchanged!
 
The foregoing scenario may be readily extended to deal with 
multiple degrees of freedom, say $N$, and---reverting to the 
original Cartesian coordinates---we find (using the summation
convention) that
\bn
  && \hskip-1cm\<p'',q''|\,e^{-i\H T}\,|p',q'\>\equiv
\lim_{\nu\ra\infty}(2\pi)^N e^{N\nu T/2}\,\int\exp\{i\tint[p_jdq^j-
h(p,q)dt]\}\,d\mu^\nu_W(p,q)\;,\\
 && |p,q\>\equiv e^{-iq^jP_j}\,e^{ip_jQ^j}|0\>\;,
\hskip1cm(Q^j+iP_j)|0\>=0\;,\hskip1cm\<0|0\>=1\;,\\
  && [Q^j,P_k]=i\delta^j_k\one\;,\\
  && \one=\tint|p,q\>\<p,q|\,d\mu_N(p,q)\;,\\
  &&\H=\tint h(p,q)\,|p,q\>\<p,q|\,d\mu_N(p,q)\;,\\
  &&d\mu_N(p,q)\equiv \Pi_{j=1}^N\,dp_j\,dq^j/2\pi\;,  
\en
and where we have used the notation $p=\{p_j\}_{j=1}^N$ and 
$q=\{q^j\}_{j=1}^N$. 
We next turn our attention to the inclusion of constraints. 
\section*{Constraints and the Projection Method}
\subsection*{Classical preliminaries}
{}From the classical point of view some of the equations of 
motion that follow from an action principle are just exactly 
{\it not} equations of motion in that they do not involve time 
derivatives but rather conditions that must be satisfied among 
the canonical variables. Consider a classical phase space action 
principle of the form
  \bn  I=\tint[p_j{\dot q}^j-h(p,q)-\l^a\phi_a(p,q)]\,dt\;, 
\en
where $\l^a=\l^a(t)$ denote Lagrange multipliers and $\phi_a(p,q)$ 
denote constraints, and $1\le a\le K\le 2N$. Stationary variation 
with respect to the dynamical variables $p_j$ and $q^j$ leads to 
the equations
 \bn  &&{\dot q}^j=\d h(p,q)/\d p_j+\l^a\d\phi_a(p,q)/\d p_j
=\{q^j,h(p,q)\}+\l^a\{q^j,\phi_a(p,q)\}\;,\\
   &&{\dot p}_j=-\d h(p,q)/\d q^j-\l^a\d\phi_a(p,q)/\d
   q^j=\{p_j,h(p,q)\}
+\l^a\{p_j,\phi_a(p,q)\}\;,  
\en
where the last version of each equation is written in terms of 
Poisson brackets.
In turn, stationary variation with respect to the Lagrange multipliers
leads to the constraint equations
  \bn  \phi_a(p,q)=0  \en
the fullfilment of which defines the {\it constraint hypersurface}. 
All processes, dynamics included, takes place on the constraint
  hypersurface. 
It follows that 
\bn
  {\dot\phi}_a(p,q)=\{\phi_a(p,q),h(p,q)\}+\l^b\{\phi_a(p,q),\phi_b(p,q)\}
=0\;. \en
Assuming that the set $\{\phi_a\}$ is a complete set of the
  constraints, 
two possible scenarios may hold. In the first scenario 
  \bn  &&\{\phi_a(p,q),\phi_b(p,q)\}=c_{ab}^{\;\;\;\;c}\phi_c(p,q)\;, \\
  &&\{\phi_a(p,q), h(p,q)\}=h_a^{\;\;b}\phi_b(p,q)\;. \en
This situation, termed {\it first class constraints}, implies that if
  the 
constraints are satisfied at one time then they will be satisfied for 
all time in the future as a consequence of the equations of motion. 
Observe in this case that the time dependence of the Lagrange
  multipliers 
$\{\l^a\}$ is not determined by these equations. To solve the
  equations 
for the variables $p_j(t)$ and $q^j(t)$ it is necessary to choose the 
Lagrange multipliers which then constitutes a ``gauge
  choice''. Nothing that is deemed physical can depend on just which 
gauge choice has been selected, and any observable, say $O(p,q)$, 
must satisfy the relation
  \bn  \{\phi_a(p,q), O(p,q)\}=o_a^{\;\;b}\phi_b(p,q)\;.  
\en
In the second situation, (30) fails, or (30) and (31) both fail, 
to hold, and as a consequence, the Lagrange multipliers must assume 
a special time dependence in order to satisfy (29). In short, 
in this case, consistency of the equations of motion determines the 
Lagrange multipliers. This case is termed {\it second class constraints}. 
Of course, one may also have a mixed case composed of some first and some 
second class constraints. In this case some of the Lagrange multipliers 
are determined while others are not.

The coefficients $c_{ab}^{\;\;\;\;c}$, $h_a^{\;\;b}$, and $o_a^{\;\;b}$ 
above may also depend on the phase space variables. However, for 
convenience, we shall restrict attention hereafter to those cases where these 
coefficients are simply constants. 
\subsection*{Quantization \`a la Dirac}
We next take up the topic of quantization of these systems. For 
the purposes of the present paper we shall confine our attention 
to the case of first class constraints.
(The case of second class 
constraints has been discussed elsewhere using the methods of the 
present paper \cite{kla2,klsh2}.)
 
According to Dirac \cite{dir2}, quantization of first class systems 
proceeds along the following line. First quantize the system as if 
there were no constraints, namely, introduce kinematical operators 
$\{P_j\}$ and $\{Q^j\}$, which fulfill (21), and a Hamiltonian 
operator $\H=\H(P,Q)$ (modulo some choice of ordering). For a general 
operator $W(P,Q)$ adopt the dynamical equation
  \bn  i{\dot W}(P,Q)=[W(P,Q),\H] \en
as usual. The constraint operators are assumed to fulfill commutation 
relations similar to (30) and (31), namely,
  \bn && [\Phi_a(P,Q),\Phi_b(P,Q)]=ic_{ab}^{\;\;\;\;c}\Phi_c(P,Q)\;,\\
     && [\Phi_a(P,Q),\H(P,Q)]=ih_a^{\;\;b}\Phi_b(P,Q)\;.  \en
Next impose the constraints to select the physical Hilbert subspace in the
form
   \bn \Phi_a(P,Q)\,|\psi\>_{\rm phy}=0\;. \en
If zero lies in the discrete spectrum of the constraint operators 
this equation offers no difficulties, and we shall content ourselves 
with that case.  On the
 other hand, if zero lies in the continuous spectrum, then some 
subtlies are involved, and one example of where such issues are 
discussed is \cite{kla2}. Observe that (34) and (35) demonstrate 
the consistency of imposing the constraints and the fact that if 
they are imposed at one time then they will hold for all subsequent 
time. This imposition of the constraints at the initial time may 
be called an {\it initial value equation}, just as in the classical theory. 

\subsection*{The projection method}
We note that the commutation relations among the constraint operators 
is that of a Lie algebra. (Indeed, including the Hamiltonian and 
noting (35), we observe that the constraints plus the Hamiltonian 
also form a Lie algebra.) For present purposes we assume that the 
group generated by this Lie algebra is compact, and we denote the 
group elements by
  \bn  e^{i\xi^a\Phi_a(P,Q)}\;.  \en
Let $\delta\xi$ denote the normalized group invariant measure, 
$\tint\delta\xi=1$, and consider the operator
 \bn \E\equiv\tint e^{i\xi^a\Phi_a(P,Q)}\,\delta\xi\;. 
\en
It is a modest exercise to establish that $\E^2=\E^\dagger=\E$, 
which are just the criteria that make $\E$ a projection operator. 
The fact that
  \bn e^{i\tau^a\Phi_a(P,Q)}\E=\tint e^{i\tau^a\Phi_a(P,Q)}
e^{i\xi^a\Phi_a(P,Q)}\delta\xi\equiv \tint e^{i(\tau{\cdot}\xi)^a
\Phi_a(P,Q)}\delta\xi=\tint e^{i\xi^a\Phi_a(P,Q)}\delta\xi=\E\;, 
\en
due simply to the invariance of the measure,
establishes that $\E$ is a projection operator onto the subspace 
where $\Phi_a=0$ for all $a$, i.e., a projection operator onto 
the physical Hilbert subspace.
We note further, based on (35), that 
  \bn e^{-i\H T}\,\E=\E\,e^{-i\H T}\,\E=\E\,e^{-i\E\H\E \,T}\,\E\;, \en

Suppose we consider a formal phase space path integral for a 
system with the classical action functional (25). The formal 
path integral reads
\bn  {\cal N}\int\exp\{i\tint[p_j{\dot q}^j-h(p,q)-\l^a\phi_a(p,q)]\,
dt\}\,\D p\,\D q
=\<p'',q''|\,{\bf T}\,e^{-i\tint[\H+\l^a\Phi_a]\,dt}\,|p',q'\>\;,  \en
which, as written, evidently depends on the choice of the Lagrange 
multipliers. Now let us impose the quantum version of the initial 
value equation, namely let us force the system at the initial time 
to lie in the physical Hilbert subspace. This we may do by 
considering the expression
 \bn \int\<p'',q''|\,{\bf T}\,e^{-i\tint[\H+\l^a\Phi_a]\,dt}\,|{\o
   p}',
{\o q}'\>\<{\o p}',{\o q}'|\E|p',q'\>\,d\mu_N({\o p}',{\o q}')\;,  
\en
which has the effect of projecting the propagator onto the physical 
subspace at time zero. Using the resolution of unity for the coherent 
states (22), it is straightforward to determine that
 \bn &&\hskip-1cm\<p'',q''|\,{\bf T}\,e^{-i\tint[\H+\l^a\Phi_a]\,dt}\,
|{\o p}',{\o q}'\>\<{\o p}',{\o q}'|\E|p',q'\>\,d\mu_N({\o p}',{\o q}')\no\\
  &&=\<p'',q''|\,{\bf T}\,e^{-i\tint[\H+\l^a\Phi_a]\,dt}\,\E\,|p',q'\>\no\\
  &&=\<p'',q''|\,e^{-i\H T}\,e^{i\tau^a\Phi_a}\,\E\,|p',q'\>\no\\
  &&=\<p'',q''|\,e^{-i\H T}\,\E\,|p',q'\>\;.  
\en
Here we have also used the properties of the Lie group relations 
(34) and (35) to separate the operators in the exponent into two 
factors, where $\{\tau^a\}$ denote parameters made of the the 
functions $\{\l^a(t)\}$ and the constants appearing in (34) and (35). 
However, whatever the choice  of the Lagrange multipliers $\{\l^a\}$, 
i.e., whatever the choice of the parameters $\{\tau^a\}$, and as 
indicated in the last line, {\it the result is completely independent 
of the Lagrange multipliers}. That is to say, the propagator projected
on the subspace spanned by $\E$ is already, and automatically, gauge 
invariant. Thus, for first class constraint systems, all that is 
necessary to achieve a gauge invariant propagator is to project 
onto the proper subspace at the initial time. 

We may introduce the projection operator $\E$ also by integrating 
over the Lagrange multipliers $\{\l^a(t)\}$ with one or another 
suitable measure. Let $C(\l)$ denote a (possibly complex) normalized 
measure, $\tint dC(\l)=1$, with the property that it introduces into 
the integrand at least one copy of the projection operator $\E$; we 
say at least one since if two (or more) are introduced the result will
be the same because $\E^2=\E$, etc. Hence, the desired propagator on 
the physical subspace may also be written as
 \bn \<p'',q''|\,e^{-i\H T}\,\E\,|p',q'\>=
 {\cal N}\int\exp\{i\tint[p_j{\dot
   q}^j-h(p,q)-\l^a\phi_a(p,q)]\,dt\}\,
\D p\,\D q\,dC(\l)\;,  \en
where, as already noted, the normalized measure $C$ is designed 
to introduce one (or more) projection operators $\E$ \cite{kla2, sha}. 
\subsubsection*{Commentary}
Readers familiar with the proposal of quantization of systems with 
first class constraints by Faddeev will note a significant difference 
in the measure for  the Lagrange multipliers. In Faddeev's treatment 
\cite{fad} the measure for the Lagrange multipliers is taken to be 
$\D\l$, namely a formally flat measure designed to introduce 
$\delta$-functionals of the classical constraints. Such a choice 
generally leads to a divergence in some of the remaining integrals 
due to a nonappearance of the variables conjugate to the constraints. 
Auxiliary conditions in the form of dynamical gauge fixing are
necessary, along with the attendant Faddeev-Popov (F-P) determinant 
needed to ensure formal canonical coordinate covariance. As is well 
known, a global choice of dynamical gauge fixing is generally
impossible, which then leads to Gribov ambiguities, and their 
associated difficulties. All these issues arose from using  a 
different measure for the Lagrange multipliers than that which is 
chosen in the present paper. In our choice the measure for the 
Lagrange multipliers is {\it normalized}, $\tint dC(\l)=1$, namely, 
an {\it average over Lagrange multipliers}, for which any divergence 
is manifestly impossible! It is of course 
true that the measure for the classical dynamical variables 
($p$ and $q$) is fixed (formally as $\D p\,\D q)$ by the requirements 
of consistency, but there is no requirement that the Lagrange
multipliers must be integrated just as the classical variables here, 
namely, from $-\infty$ to $\infty$ with a flat weighting. Who says we 
must enforce the {\it classical} constraints when doing the {\it
  quantum} theory? No one of course, and in a general sense that 
is the only freedom that has been used to avoid dynamical gauge 
fixing, F-P determinants, Gribov ambiguities, ghosts, ghosts of
ghosts, etc., and all the other machinery of the BRST and BFV 
formalisms \cite{hen, grib}. Our results involve only the original 
phase space variables as augmented by the Lagrange multipliers, 
do not in any way entail enlargement or even in many cases a reduction
in the number of the original variables,  lead to results that are 
entirely gauge invariant and satisfactory, and at the same time avoid 
altogether the whole galaxy of issues listed above which are
instigated 
by using a flat measure for the Lagrange multipliers. 
\section*{Yang-Mills Type Gauge Models}
In this section we illustrate the imposition of constraints for a 
special class of models which we call Yang-Mills type \cite{klsh}. 
In the classical theory, in which we assume we have chosen 
Cartesian coordinates in phase space, the constraints are taken as
  \bn  \phi_a(p,q)=p_a+A_{ab}^{\;\;\;\;c}q^bp_c\;, 
\en
where the parameters $A_{ab}^{\;\;\;\;c}$ are antisymmetric 
in the indices $b,c$. This constraint induces a shift in the 
$a$th coordinate  and a rotation in the $b-c$ plane. Such 
constraints are broad enough to cover the usual Yang-Mills 
theories (where $p$ plays the role of the electrical field strength 
$E$ and $q$ the role of the vector potential $A$). 
It is clear that such constraints commute among themselves to form 
a Lie algebra. In the quantum theory the constraint operators are taken to be 
  \bn \Phi_a(P,Q)=P_a+A_{ab}^{\;\;\;\;c}Q^bP_c\;,  \en
and it is an attractive feature of such constraints that
  \bn e^{i\Omega^a\Phi_a}\,|p,q\>=|p^\Omega,q^\Omega\>\;, \en
namely that the unitary transformation generated by the 
constraints takes one coherent state into another coherent state. Here
 \bn  p^\Omega\equiv e^{-\Omega^a\,{\rm ad}\,\phi_a}p\;,\hskip1cm 
q^\Omega\equiv e^{-\Omega^a\,{\rm ad}\,\phi_a}q\;, \en
where ${\rm ad}\,\phi_a(\cdot)\equiv \{\phi_a,\,\cdot\,\}$ and the 
exponential is defined by its power series expansion. 

The inclusion of such constraints into a dynamical system is quite 
straightforward. For that goal we first note that
 \bn \tint \<p'',q''|\,e^{-i\H
T}\,e^{i\Omega^a\Phi_a}\,|p',q'\>\,\delta\Omega=
  \tint\<p'',q''|\,e^{-i\H T}\,|p'^{\,\Omega},q'^{\,\Omega}\>\,\delta\Omega=
   \<p'',q''|\,e^{-i\H T}\,\E\,|p',q'\>\;, \en
which asserts that in order to insert the desired projection operator 
it is only necessary to average the initial coherent state labels over
the gauge transformations they experience. In turn this means that
\bn \<p'',q''|\,e^{-i\H T}\,\E\,|p',q'\>=
   \lim_{\nu\ra\infty}(2\pi)^Ne^{N\nu
     T/2}\int\exp\{i\tint[pdq-h(p,q)\,
dt]\}\,d\mu^\nu_W(p,q)\,\delta\Omega\;,  \en
where it is understood that the Wiener paths are pinned initially at 
$p^{\,\Omega}$ and $q^{\,\Omega}$. We can make this expression appear 
more familiar by making a change of variables within the well defined 
path integral. In particular, we let \bn p(t)\ra e^{\tint_t^Tds\,
\omega^a(s)\,{\rm ad}\,\phi_a}\,p(t)\;,\hskip1cm  q(t)\ra
e^{\tint_t^Tds\,
\omega^a(s)\,{\rm ad}\,\phi_a}\,q(t)\;, \en
where the functions $\omega^a(s)$ are arbitrary save for the condition that
 \bn \int_0^T \omega^a(s)\,ds=\Omega^a\;. \en
Since $(p'^{\,\Omega})^{-\Omega}\equiv p'$ and
$(q'{\,^\Omega})^{-\Omega}
\equiv q'$, this transformation has the effect of removing any
influence 
of the gauge transformation on the initial labels $p',q'$, and instead
redistributing that influence throughout the time evolution of the
path integral. As a proper coordinate transformation within a well 
defined path integral, we can readily determine the effect of such a 
change of variables. In particular, appealing to a formal notation for
clarity, we find after such a variable change that (50) becomes
 \bn&&\hskip-.6cm\<p'',q''|\,e^{-i\H
   T}\,\E\,|p',q'\>=\lim_{\nu\ra\infty}
{\cal N}\int\exp\{i\tint[p_j{\dot
  q}^j-\omega^a\phi_a(p,q)-h(p,q)]\,dt\}
\no\\  &&\times\exp\{-(1/2\nu)\tint[({\dot p}-\omega^a\{\phi_a,p\})^2+
({\dot q}-\omega^a\{\phi_a,q\})^2]\,dt\}\,\D p\,\D q\,\delta\Omega \;.  
\en
In this expression a ``new'' term has appeared in the classical action that
looks like a sum of Lagrange multipliers times the constraints, and 
drift terms have arisen in the Wiener measure regularization. Note 
also what this formula states: On the right side is a path integral 
which superficially depends on the functions $\{\omega^a(s)\}$, $0\le 
s\le T$. On the left side, there is no such dependence. In other
words, 
although the path integral {\it appears} to depend on $\{\omega^a\}$, 
in fact it does not. Therefore we are free to {\it average} the 
right-hand side of (53) over the functions $\{\omega\}$ and still 
obtain the desired answer. Let the measure $C(\omega)$ denote such a 
measure chosen so as to include the initial average over the 
variables $\Omega$ as well, and normalized so that $\tint dC(\omega)=1$. 
The only requirement we impose on this measure is that it introduce, 
as did the original measure over the variables $\Omega$, at least one 
projection operator $\E$. In this case we find the important phase 
space path integral representation given by
 \bn&&\hskip-.6cm\<p'',q''|\,e^{-i\H T}\,\E\,|p',q'\>\no\\
&&=\lim_{\nu\ra\infty}{\cal N}\int\exp\{i\tint[p_j{\dot q}^j-\omega^a
\phi_a(p,q)-h(p,q)]\,dt\}\no\\  &&\times\exp\{-(1/2\nu)\tint[({\dot
  p}-\omega^a\{\phi_a,p\})^2+({\dot
  q}-\omega^a\{\phi_a,q\})^2]\,dt\}\,
\D p\,\D q\,dC(\omega) \;.  \en
Here we can really see the variability of the Lagrange multipliers in 
the path integration and how the proper choice of measure for them 
can lead to the desired gauge invariant result without any 
additional complications. 

The only topic left to discuss is what should we take for the measure 
$C$. In fact there are many answers to that question, but, for
brevity, we shall only indicate one of them. We suppose that 
our compact gauge group is semisimple and therefore admits a 
group-induced, positive definite metric $g_{ab}(\omega)$. Given 
that metric on the group manifold we introduce a Wiener measure 
formally given by
 \bn dC(\omega)= {\cal M}\exp[-\half\tint g_{ab}\,{\dot\omega}^a
{\dot\omega}^b\,dt]\,\Pi_t\delta\omega(t)\;, \en
where this measure is {\it not} pinned either at $t=0$ or at $t=T$. 
A normalized measure without pinning is made possible because the 
space of variables $\{\omega^a\}$ at any one time is compact. The 
formal constant $\cal M$ is chosen to ensure $\tint dC(\omega)=1$.

In summary, we have illustrated how the use of coherent states and a natural 
flat phase space metric can be used to develop a coordinate-free 
quantization procedure for systems without constraints as well as for 
systems with constraints. We hope this introductory paper may encourage 
the reader to delve further into this fascinating subject.

\section*{Dedication}
It is a pleasure to dedicate this paper to the 65th birthday of 
Hiroshi Ezawa, which was the main event celebrated at the 2nd 
Jagna Workshop. Over a number of years, one of the authors (J.R.K.) 
has enjoyed numerous interactions with the honoree including, but 
not limited to, two years of close collaboration at Bell Laboratories, 
and several visits to Tokyo to share in the scientific and social 
life of Japan. Such great personal interactions and experiences are 
truly what makes life worthwhile!

\section*{Acknowlegements}
It is a great pleasure to thank the organizers Chris and Victoria 
Bernido, and their extended families, for hosting such a pleasant 
meeting, and which additionally offered the participants a delightful 
glimpse of Philippine country life. It was an additional pleasure to 
meet old friends and to make new ones among the local participants. 
We hope the series of Jagna workshops will continue for many years to come!

\end{document}